\documentclass[aps,pra,twocolumn,showpacs]{revtex4}
\usepackage{amsmath, amssymb}
\usepackage{color, graphicx}

\begin{document}

\title{Sub-binomial light}
\author{J. Sperling} \email{jan.sperling2@uni-rostock.de}\affiliation{Arbeitsgruppe Quantenoptik, Institut f\"ur Physik, Universit\"at Rostock, D-18051 Rostock, Germany}
\author{W. Vogel}\affiliation{Arbeitsgruppe Quantenoptik, Institut f\"ur Physik, Universit\"at Rostock, D-18051 Rostock, Germany}
\author{G. S.  Agarwal}\affiliation{Department of Physics, Oklahoma State University, Stillwater, OK, USA}
\pacs{42.50.Ar, 03.65.Wj, 42.50.-p}
\date{\today}

\begin{abstract}
	The click statistics from on-off detector systems is quite different from the counting statistics of the more traditional detectors.
	This necessitates introduction of new parameters to characterize the nonclassicality of fields from measurements using on-off detectors.
	To properly replace the Mandel $Q_{\rm M}$ parameter, we introduce a parameter $Q_{\rm B}$.
	A negative value represents a sub-binomial statistics.
	This is possible only for quantum fields, even for super-Poisson light.
	It eliminates the problems encountered in discerning nonclassicality using Mandel's $Q_{\rm M}$ for on-off data.
\end{abstract}

\maketitle

\section{Introduction}
	Nonclassicality of the radiation fields has been at the heart of Quantum Optics.
	One uses detectors which work by absorption of photons and hence normally ordered correlations are the ones measured directly~\cite{VogelBook,MandelWolf,VogelReview}.
	Nonclassicality in Quantum Optics has been therefore formulated in terms of the nonclassical properties of the $P$~function associated with the density matrix of the quantum fields~\cite{GSfunction1,GSfunction2}.
	However, the $P$~function itself is not directly measurable.
	Mandel introduced an experimentally deducible measure of nonclassicality, namely the $Q_{\rm M}$ parameter, defined by
	\begin{align}\label{Eq:Q}
		Q_{\rm M}=\frac{\langle (\Delta n)^2\rangle}{\langle n\rangle}-1, 
	\end{align}
	with $\langle n\rangle$ and $\langle (\Delta n)^2\rangle$ being the classical mean value and the classical variance of the photoelectric statistics~\cite{QMandel}.
	If $Q_{\rm M}$ is negative, then the photo counting statistics is of sub-Poisson type and we conclude that the field is nonclassical. The very first experimental demonstration of this nonclassical effect was given in~\cite{Short}.

	More recently it has become necessary to use photon number resolving (PNR) detectors to discriminate between states with definite photon numbers~\cite{Yamamoto,DetArray,TMD1,TMD2}.
        Since such detectors are not directly available, one uses on-off detector systems (avalanche photodiodes)~\cite{SinglePhoton}.
	Such detector systems have been characterized by tomographic methods~\cite{
        DetectorTomography2}.
	The deduction of nonclassicality using measurements with avalanche photodiodes and using the Mandel $Q_{\rm M}$ parameter meets with difficulties.
	For example, even if the field is completely classical, then $Q_{\rm M}$ can be negative~\cite{SpeVoA}.

	In this article we present a solution to this difficulty by introducing an appropriate measure of nonclassicality using the data from on-off detector systems.
	The condition $Q_{\rm B}<0$ characterizes the sub-binomial click statistics of light. 
	We provide a physical justification for the new measure and we show by several examples the validity of the binomial $Q_{\rm B}$ parameter.
	Whenever the discrimination of adjacent photon numbers is of relevance for applications in modern quantum technologies, the notion of sub-binomial light is expected to play a vital role.


\section{The binomial $Q_{\rm B}$ parameter}
	The traditional detectors work on the principle that a photo electron is emitted if a photon is absorbed.
	Perturbation theory shows that the emission probability is proportional to the intensity of light and this leads to the counting distribution~\cite{QMandel,KK},
	\begin{align}\label{Eq:TrueStat}
		p_n=\langle {:} \frac{(\eta\hat n+\nu)^n}{n!}e^{-(\eta\hat n+\nu)} {:}\rangle.
	\end{align}
        Herein, the operator $\hat n$ represents the photon number, $\eta$ the detection efficiency, $\nu$ the number of noise or dark counts, and the $:\,\cdot\,:$ notation indicates the normal ordering prescription.
	For on-off detectors the mechanism is different.
	The detector clicks for any number of photons and does not click if the field is in the vacuum state.
	We showed recently that if one employs $N$ on-off detectors, then the counting distribution is given by~\cite{SpeVoA}
	\begin{align}\label{Eq:PNRcounting}
		c_k=\langle {:} \frac{N!}{k!(N-k)!}\left(e^{-\frac{\eta\hat n+\nu}{N}}\right)^{N-k}\left(\hat 1-e^{-\frac{(\eta\hat n+\nu)}{N}}\right)^{k} {:}\rangle.
	\end{align}
	We also observed, that the counting or click statistics converges to the true statistics with increase in the number of detectors.
	However this convergence is slow as it goes as $1/N$.
	For coherent states, Eqs.~(\ref{Eq:TrueStat})~and~(\ref{Eq:PNRcounting}) reduce to the Poisson and binomial statistics, respectively.

	Since the counting distribution has a different form, one would expect that one needs a measure different from the Mandel $Q_{\rm M}$ parameter to characterize nonclassicality.
	Note that the traditional photo counting distribution involves the expectation of a normally ordered Poisson distribution, whereas the click statistics involves a normally ordered binomial one.
	We expect that an appropriate measure of the nonclassical statistics would be the {\em sub-binomiality} of the distribution.
	Hence we introduce $Q_{\rm B}$ defined by
	\begin{align}\label{Eq:QN}
		Q_{\rm B}=N\frac{\langle (\Delta c)^2 \rangle}{\langle c \rangle(N-\langle c \rangle)}-1,
	\end{align}
	where $\langle c \rangle$ is the mean number of clicks, and $\langle (\Delta c)^2 \rangle$ the variance of the click statistics $(c_k)_{k=0}^N$,
	\begin{align}\label{Eq:MeanVar}
		\langle c \rangle=\sum_{k=0}^N k\,c_k
		\quad\text{and}\quad
		\langle (\Delta c)^2 \rangle=\sum_{k=0}^N (k-\langle c \rangle)^2 c_k.
	\end{align}
        The moments $\langle c \rangle$ and $\langle (\Delta c)^2 \rangle$ are defined the sense of classical probabilistic quantities.

	This $Q_{\rm B}$ possesses the properties:
	\begin{enumerate}
		\item The $Q_{\rm B}$ parameter must not yield negative values for classical states;
		\item For any quantum state having a binomial counting statistics $(c_k)_{k=0}^N$, $Q_{\rm B}$ should be zero;
		\item It is based on first and second moments of $(c_k)_{k=0}^N$;
		\item For $N\to\infty$, $Q_{\rm B}$ should converge to $Q_{\rm M}$.
	\end{enumerate}

	The definition of $Q_{\rm B}$ requires at least two on-off detectors.
	A single on-off detector only yields one click and no click with probability  $c_0=p$ and $c_1=1-p$ ($0\leq p\leq1$), respectively.
	Thus, any quantum state has a binomial statistics, $Q_{\rm B}=0$, as long as a single detector is used.

	We next prove that $Q_{\rm B}<0$ is a measure of nonclassicality for measurements with $N$ on-off detectors.
	For this purpose, it is convenient to use the generating function of the click statistics
	\begin{align}
		f(x)=&\sum_{k=0}^N c_k\,x^k
		=\langle{:}\left[x\left(\hat 1-e^{-\frac{\eta\hat n+\nu}{N}}\right)+e^{-\frac{\eta\hat n+\nu}{N}}\right]^N{:}\rangle.
	\end{align}
	From the derivatives of $f$, one can obtain all moments of the statistics.
	It can be shown, that the variance $\langle(\Delta c)^2\rangle$ reads as
	\begin{align}
		\nonumber \langle (\Delta c)^2\rangle=& N(N-1)\langle{:}\left(\Delta e^{-\frac{\eta\hat n+\nu}{N}}\right)^2{:}\rangle
		\\&+N\left(1-\frac{\langle c\rangle}{N}\right)\frac{\langle c\rangle}{N},
	\end{align}
	with  $\langle c\rangle/N=1-\langle{:}\exp[-(\eta\hat n+\nu)/N]{:}\rangle$, cf.~Appendix~A.
	Rewriting this equation according to Eq.~(\ref{Eq:QN}),
	\begin{align}
		Q_{\rm B} =\frac{(N-1)\langle{:}\left(\Delta e^{-\frac{\eta\hat n+\nu}{N}}\right)^2{:}\rangle}{\langle{:} e^{-\frac{\eta\hat n+\nu}{N}}{:}\rangle\left(1-\langle{:} e^{-\frac{\eta\hat n+\nu}{N}}{:}\rangle\right)},\label{Eq:NormOrdQN}
	\end{align}
	we obtain the binomial $Q_{\rm B}$ parameter in its explicite form. 
	Note that the parameter $Q_{\rm B}$, as it is clearly seen from this result, depends on higher-order moments of the photon number statistics, which is not the case for Mandel $Q_{\rm M}$ parameter given in Eq.~(\ref{Eq:Q}).
	Such higher order moments are beginning to be studied in experiments~\cite{HighCorr1,HighCorr2}.
	For a classical state, when the $P$-function has the properties of a classical probability distribution~\cite{NonClDef}, it yields that any normally ordered variance is non-negative.
	In addition, $\langle c\rangle/N$ and $1-\langle c\rangle/N$ are non-negative mean values.
	It follows for classical states
	\begin{align}
		Q_{\rm B}\geq 0.
	\end{align}
	Let us note that the individual expectation values are independent of the phase.
	Altogether, this proves the claim that a negative binomial $Q_{\rm B}$ value implies a nonclassical photon statistics.

	Concerning the convergence properties of $Q_{\rm B}$, we can use the following result.
	In Ref.~\cite{SpeVoA}, we have already shown, that the click statistics converges to the photo statistics for $N\to\infty$.
	It follows that $\langle (\Delta c)^2 \rangle/\langle  c \rangle$ converges to $\langle (\Delta n)^2 \rangle/\langle n \rangle$.
	The only difference left between $Q_{\rm M}$ and $Q_{\rm B}$, cf. Eqs.~(\ref{Eq:Q})~and~(\ref{Eq:QN}), is
	\begin{align}
		\frac{N}{N-\langle c\rangle}=\frac{1}{1-\frac{\langle c\rangle}{N}}.
	\end{align}
	Since, $\langle c\rangle$ converges to the finite value of $\langle n\rangle$, we obtain
        \begin{align}
		Q_{\rm B}\to Q_{\rm M}\quad\text{for}\quad N\to\infty.
	\end{align}

	As  a last property we verify that for coherent states $Q_{\rm B}=0$.
	According to Eq.~(\ref{Eq:PNRcounting}), $(c_k)_{k=1}^N$ is a binomial distribution with
	\begin{align}
		\langle c\rangle=&N\left(1-e^{-\frac{\eta|\alpha|^2+\nu}{N}}\right),\\
		\nonumber \langle (\Delta c)^2\rangle=&N\left(1-e^{-\frac{\eta|\alpha|^2+\nu}{N}}\right)e^{-\frac{\eta|\alpha|^2+\nu}{N}}.
	\end{align}
	Applying the binomial $Q_{\rm B}$ parameter, we obtain the desired interpretation, $Q_{\rm B}=0$.
	The parameter $Q_{\rm B}$ does not lead to fake nonclassicality, for any choice of noise or detection efficiency.
	For more general states having a binomial statistics, we can formulate similarly $Q_{\rm B}=0$.

	Our binomial $Q_{\rm B}$ parameter is directly constructed for measurements with on-off detector systems, including imperfections.
	It can discern nonclassicality in experiments using only two or more on-off detectors.
	In the following, we apply the $Q_{\rm B}$ parameter to typical examples in Quantum Optics.
        We consider three kinds of states having comparable mean photon numbers.
	Usually the main source of imperfections is caused by the quantum efficiency $\eta<1$, so that we may assume a negligible noise count rate, $\nu\approx 0$.


\section{Thermal states}
	First, we may consider a classical, thermal state, with a mean photon number $\langle n\rangle=\bar n$ and a variance $\langle (\Delta n)^2\rangle=\bar n(\bar n+1)$.
	Such a state enables us to highlight the difficulty associated with the value of $Q_{\rm M}$ for data from on-off detectors.
	It has a positive Mandel parameter, $Q_{\rm M}=\bar n$.
	The efficiency simply scales the mean photon number $\bar n$ to a smaller value, $\eta\bar n$.

	Using the $P$~function of the thermal state, we obtain the click statistics in the form of a beta-binomial distribution $c_k=f_{N,\alpha,\beta}(k)$ for $\alpha=1$ and $\beta=N/{\bar n}$,
	\begin{align}
		 c_k=&\frac{N!}{k!(N-k)!}\frac{\Gamma\left(N-k+\frac{N}{\bar n}\right)\Gamma\left(k+1\right)}{\Gamma\left(N+1+\frac{N}{\bar n}\right)}\frac{N}{\bar n},
	\end{align}
	see~Appendix~B~1.
	The well-known mean values and variances of such distributions yield $Q_{\rm B}$ values, 
	\begin{align}
		Q_{\rm B}=\frac{N+\frac{N}{\bar n}+1}{\frac{N}{\bar n}+2}-1=\frac{N-1}{\frac{N}{\bar n}+2}.
	\end{align}

\begin{center}
	\begin{figure}[ht]
		\includegraphics*[width=5.4cm]{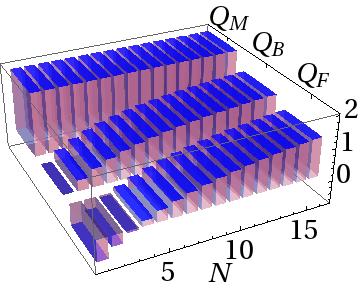}
		\caption{(color online)
		Different $Q$ parameters are plotted for a classical, thermal state, with $\langle n\rangle=\eta\bar n=2$ and $\nu=0$.
		The true Mandel parameter is $Q_{\rm M}$, the binomial one is $Q_{\rm B}$.
		Using the definition in Mandel's form for click statistics yields $Q_{\rm F}$.
		}
	\label{Fig:FalseQPoisson}
	\end{figure}
	\end{center}

	In Fig.~\ref{Fig:FalseQPoisson}, we plotted different parameters depending on the number $N$ of on-off detectors, with $2\le N\leq 16$.
	Determining the Mandel parameter value from the click statistics -- denoted as $Q_{\rm F}$ --  leads to fake nonclassicality.
	The binomial parameter, $Q_{\rm B}>0$, correctly displays the classicality of thermal light.
	For large numbers of on-off detectors, $Q_{\rm B}$ approaches the value of $Q_{\rm M}=\bar n$.

\section{Fock states}
	Second, we may study a Fock state for $m$ photons.
	To solely consider the effects of on-off detectors, we choose the detection efficiency $\eta=1$.
	The photon statistics of the Fock state is a singular one, $p_n=\delta_{n,m}$ and $Q_{\rm M}=-1$.

        To obtain the click statistics from the true photo statistics, we apply Eq.~(14) of Ref.~\cite{SpeVoA} to get
	\begin{align}
		c_k=\sum_{n=0}^\infty \frac{N!}{k!(N-k)!}\frac{\partial_y^n\left(e^y-1\right)^k|_{y=0}}{N^n}\, p_n.
	\end{align}
	The particular example of a Fock state yields
	\begin{align}
		c_k=\frac{N!}{k!(N-k)!}\frac{\partial_y^m\left(e^y-1\right)^k|_{y=0}}{N^m}.
	\end{align}
	The mean click number $\langle c\rangle$ and the variance $\langle(\Delta c)^2\rangle$ are given by 
	\begin{align}
		\langle c\rangle=&N\left(1-\left[1-\frac{1}{N}\right]^m\right),\\
		\nonumber\langle (\Delta c)^2\rangle=&N(N-1)\left(1-2\left[1-\frac{1}{N}\right]^m+\left[1-\frac{2}{N}\right]^m\right)\\
		&+\langle c\rangle-\langle c\rangle^2,
	\end{align}
        cf.~Appendix~B~2.
	Thus, the analytical expression of the $Q_{\rm B}$ parameter for $m$ photons is
	\begin{align}
		Q_{\rm B}=(N-1)\frac{N^m(N-2)^m-(N-1)^{2m}}{(N^m-(N-1)^m)(N-1)^m}.
	\end{align}

\begin{center}
	\begin{figure}[ht]
		\includegraphics*[width=5.4cm]{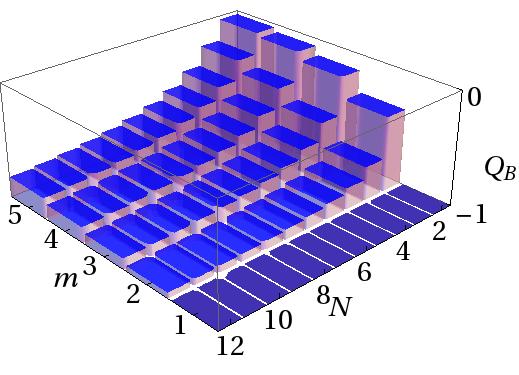}
		\caption{(color online)
		The plot shows the binomial $Q_{\rm B}$ parameter for the click statistics measured by $N$ on/off detectors, with $2\leq N\leq 12$.
		The computed example shows the sub-binomial statistics of $m$ photon Fock states ($1\leq m\leq 5$).
		}
	\label{Fig:TrueQSubBinominialN2}
	\end{figure}
	\end{center}

	In Fig.~\ref{Fig:TrueQSubBinominialN2}, we plotted the $Q_{\rm B}$ parameter depending on the number of photons, $m$, and the number of available on-off detectors, $N$.
	The verification of a nonclassical photon number statistics can be directly observed from $Q_{\rm B}<0$.
	This is possible, although the considered detector system is unable to measure the true photo statistics.
	It is also clear, that a larger number of photons $m$ requires a higher number of detectors to significantly identify nonclassicality.
	Surprisingly, a measurement using only two on-off detectors can be used to infer nonclassical light.
	We can also observe that for large numbers of on-off detectors $Q_{\rm B}$ approaches the value $Q_{\rm M}=-1$.

\section{Single-photon-added thermal state}
	In this last example, we show that the binomial $Q_{\rm B}$ parameter can detect nonclassical photon statistics beyond sub-Poisson ones.
	For this purpose let us study a  single-photon-added thermal state (SPATS)~\cite{SPATS1}.
	This state has been experimentally realized~\cite{SPATS2}, and its nonclassicality has been verified by reconstructing its $P$~function~\cite{SPATS3}.
	The $P$~function of the SPATS is given by
	\begin{align}
		P_{\rm SPATS}(\alpha)=\frac{1}{\pi\bar n^3}\left[(1+\bar n)|\alpha|^2-\bar n\right]e^{-\frac{|\alpha|^2}{\bar n}}.
	\end{align}
	A straightforward computation of $Q_{\rm M}$ yields for the efficiency $\eta$
	\begin{align}
		Q_{\rm M}=\eta\frac{\bar n^2-\frac{1}{2}}{\bar n+\frac{1}{2}}
		\left\lbrace\begin{array}{ccc}
			\geq 0 &\quad\text{for}\quad& \bar n\geq \sqrt{0.5},\\
			< 0 &\quad\text{for}\quad& \bar n< \sqrt{0.5}.
		\end{array}\right.
	\end{align}
	For $\bar n>\sqrt{1/2}$, the $Q_{\rm M}$ parameter cannot identify the nonclassicality of the SPATS.

	In the following we apply the $Q_{\rm B}$ parameter to the SPATS.
	Some algebra, using the $P_{\rm SPATS}$~distribution, yields 
	\begin{align}
		Q_{\rm B}=(N-1)\frac{I\left(\frac{2\eta}{N}\right)-I\left(\frac{\eta}{N}\right)^2}{I\left(\frac{\eta}{N}\right)\left[1-I\left(\frac{\eta}{N}\right)\right]},
	\end{align}
        the needed integral $I(\lambda)$ can be analytically computed, see~Appendix~B~3.
	In Fig.~\ref{Fig:SPATS}, we plotted the binomial $Q_{\rm B}$ parameter depending on $\bar n$.
	For the chosen parameters we have a super-Poisson statistics, $Q_{\rm M}\geq0$.
	However, we find regions with a sub-binomial statistics, $Q_{\rm B}<0$.
	We obtain the surprising result, that the click statistics can be more suitable to detect nonclassicality than the $Q_{\rm M}$ parameter, even for a small number of on-off detectors.
	\begin{center}
	\begin{figure}[ht]
		\includegraphics*[width=4cm]{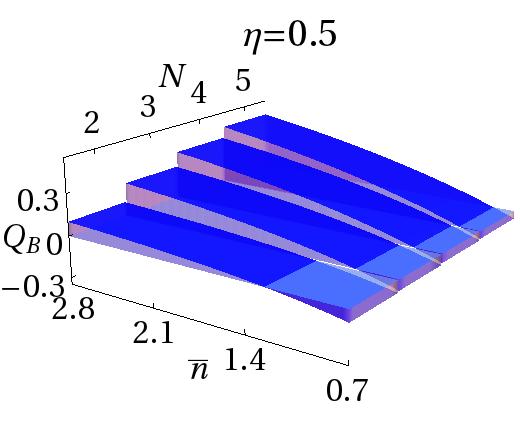}\quad
		\includegraphics*[width=4cm]{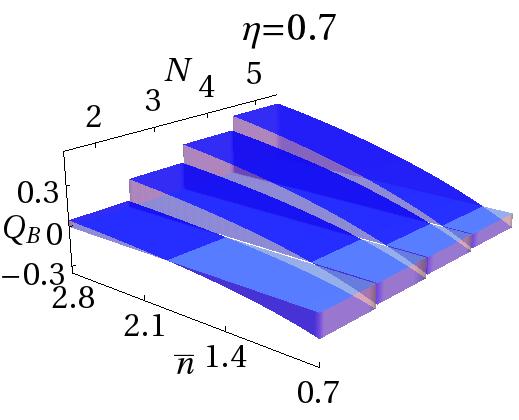}\\
		\includegraphics*[width=4cm]{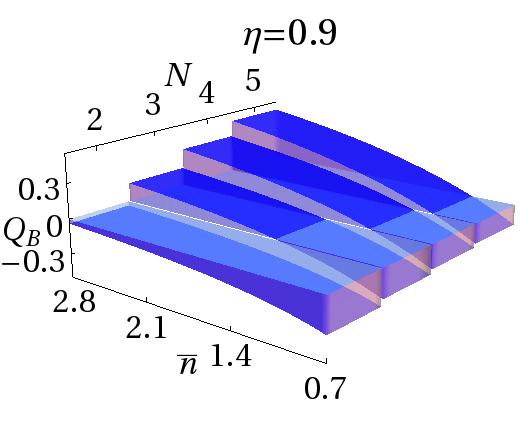}
		\caption{(color online)
		The plot shows the binomial $Q_{\rm B}$ parameter for the counting statistics measured by a PNR detector with $N=2,3,4,5$ on/off detectors.
		The computed example is a SPATS with a mean thermal photon number $\sqrt{0.5}\leq\bar n\leq 4\sqrt{0.5}$.
		The individual plots have a quantum efficiency $\eta=0.5,0.7,0.9$.
	}\label{Fig:SPATS}
	\end{figure}
	\end{center}


\section{Summary and Conclusions}
	We established the binomial $Q_{\rm B}$ parameter.
	It serves for the identification of nonclassical radiation measured with multiple on-off detectors including imperfections.
	A negative parameter, $Q_{\rm B}<0$, refers to as sub-binomial light.
	We showed that the binomial $Q_{\rm B}$ parameter convergences to the original Mandel $Q_{\rm M}$ parameter for large numbers of on-off detectors.
	It is worth mentioning that our method does not require a reconstruction of the true photon number statistics.

	We applied our method to typical states, for example, Fock states representing sub-binomial light.
	We also studied the statistics of a single-photon-added thermal state measured by only a small number of on-off detectors.
	In this case the $Q_{\rm B}$ parameter can identify nonclassical, in particular sub-binomial, states of light beyond the $Q_{\rm M}$ parameter.
	From a more general perspective, sub-binomial light and its characterization may become of vital relevance whenever modern quantum techologies require the discrimination of photon number states.

\section*{Acknowledgment}
	This work was supported by the Deutsche Forschungsgemeinschaft through SFB 652.


\appendix
\begin{widetext}
	\section{Derrivation of $Q_{\rm B}$}\label{App:Derive}
		Let us derive the $Q_{\rm B}$ parameter.
		For this reason, we consider the generating function of the click statistics as
		\begin{align}
			\label{Eq:GenFct}
			f(x)=\sum_{k=0}^N c_k x^k=\langle{:}\left[x\left(\hat 1-e^{-\frac{\eta\hat n+\nu}{N}}\right)+e^{-\frac{\eta\hat n+\nu}{N}} \right]^N{:}\rangle
		\end{align}
		which directly follows from the click statistics $c_k=\frac{N!}{k!(N-k)!}\langle{:}\left(e^{-\frac{\eta\hat n+\nu}{N}}\right)^{k}\left(\hat 1-e^{-\frac{\eta\hat n+\nu}{N}}\right)^{N-k}{:}\rangle$.
		Now we identify
		\begin{align}
			\label{Eq:ExpGen} \left.\partial_xf(x)\right|_{x=1}&=\sum_k k\,c_k=\langle c\rangle
			=N\langle{:}\left(\hat 1-e^{-\frac{\eta\hat n+\nu}{N}}\right){:}\rangle,\\
			\left.\partial_x^2f(x)\right|_{x=1}&=\sum_k k(k-1)\,c_k=\langle c^2\rangle-\langle c\rangle
			=N(N-1)\langle{:}\left(\hat 1-e^{-\frac{\eta\hat n+\nu}{N}}\right)^2{:}\rangle.
		\end{align}
		We may write the click variance $\langle(\Delta c)^2\rangle$ as
		\begin{align}
			\nonumber \langle (\Delta c)^2\rangle=&\sum_{k=0}^N (k-\langle c\rangle)^2 c_k=\langle c^2\rangle-\langle c\rangle^2\\
			\nonumber =& N(N-1)\langle{:}\left(\hat 1-e^{-\frac{\eta\hat n+\nu}{N}}\right)^2{:}\rangle+N\langle{:}\left(\hat 1-e^{-\frac{\eta\hat n+\nu}{N}}\right){:}\rangle-N^2\langle{:}\left(\hat 1-e^{-\frac{\eta\hat n+\nu}{N}}\right){:}\rangle^2\\
			\nonumber =& N(N-1)\langle{:}\Delta\left(\hat 1-e^{-\frac{\eta\hat n+\nu}{N}}\right)^2{:}\rangle
			+N\left(1-\langle{:}e^{-\frac{\eta\hat n+\nu}{N}}{:}\rangle\right)\langle{:}e^{-\frac{\eta\hat n+\nu}{N}}{:}\rangle\\
			=& N(N-1)\langle{:}\left(\Delta e^{-\frac{\eta\hat n+\nu}{N}}\right)^2{:}\rangle
			+N\left(1-\frac{\langle c\rangle}{N}\right)\frac{\langle c\rangle}{N},
		\end{align}
		using the fact that the variance of a quantity $\hat 1-\hat L$ is the same as for $\hat L$.
		For a classical state, which means that $P$ has the properties of a classical probability distribution, any normally ordered variance is non-negative, $\langle{:}(\Delta \hat L)^2{:}\rangle\geq 0$.
		In addition, the expectation value as defined in Eq.~(\ref{Eq:ExpGen}) is a value between $0$ and $N$.
		It follows for classical states that
		\begin{align}\label{Eq:QB}
			Q_{\rm B}=\frac{N(N-1)\langle{:}\left(\Delta e^{-\frac{\eta\hat n+\nu}{N}}\right)^2{:}\rangle}{N\left(1-\frac{\langle c\rangle}{N}\right)\frac{\langle c\rangle}{N}}
			=\frac{(N-1)\langle{:}\left(\Delta e^{-\frac{\eta\hat n+\nu}{N}}\right)^2{:}\rangle}{\langle{:} e^{-\frac{\eta\hat n+\nu}{N}}{:}\rangle\left(1-\langle{:} e^{-\frac{\eta\hat n+\nu}{N}}{:}\rangle\right)}\geq 0,
		\end{align}

		Let us note that this derivation can be similarily obtained for the Mandel $Q_{\rm M}$ parameter.
		Its statistics yields a generating function
		\begin{align}
			f(x)=\sum_{n\in\mathbb N} p_n x^n=\langle{:}e^{(x-1)\hat n}{:}\rangle,
		\end{align}
		with the simplification $\eta=1$ and $\nu=0$.
		In analogy to the above considerations we find that
		\begin{align}
			\langle(\Delta n)^2\rangle=\langle{:}(\Delta \hat n)^2{:}\rangle+\langle n\rangle.
		\end{align}
		This equation can be reformulated as
		\begin{align}
			\frac{\langle(\Delta n)^2\rangle}{\langle n\rangle}-1=\frac{\langle{:}(\Delta \hat n)^2{:}\rangle}{\langle n\rangle}=Q_{\rm M}.
		\end{align}

	\section{Examples}\label{App:Examples}

	\subsection{Thermal states}
		To determine the properties of the thermal state, we use its $P$~function.
		From $P(\alpha)=\exp[-|\alpha|^2/\bar n]/(\pi\bar n)$, we obtain the counting statistics as
		\begin{align}
			\nonumber c_k=&\frac{N!}{k!(N-k)!}\frac{2}{\bar n}\int_0^\infty dr\,r \left(e^{-\frac{r^2}{N}}\right)^{N-k+\frac{N}{\bar n}} \left(1-e^{-\frac{r^2}{N}}\right)^k
			=\frac{N!}{k!(N-k)!}\frac{N}{\bar n}\int_0^1 ds\, \left(s\right)^{N-k+\frac{N}{\bar n}-1} \left(1-s\right)^k\\
			\nonumber =&\frac{N!}{k!(N-k)!}\frac{\Gamma\left(N-k+\frac{N}{\bar n}\right)\Gamma\left(k+1\right)}{\Gamma\left(N+1+\frac{N}{\bar n}\right)}\frac{N}{\bar n}=\frac{\Gamma\left(N+1\right)}{\Gamma\left(k+1\right)\Gamma\left(N-k+1\right)}
			\frac{\Gamma\left(\alpha+k\right)\Gamma\left(N-k+\beta\right)}{\Gamma\left(\alpha+\beta+N\right)}
			\frac{\Gamma\left(\alpha+\beta\right)}{\Gamma\left(\alpha\right)\Gamma\left(\beta\right)}\\
			=&f_{N,\alpha,\beta}(k),
		\end{align}
		being the beta-binomial distribution $f_{N,\alpha,\beta}$ with $\alpha=1$ and $\beta=N/{\bar n}$.
		This well-known distribution has a mean value of $\langle c\rangle=N/(\alpha+\beta)$ and a variance of
		\begin{align}
			\langle(\Delta c)^2\rangle=\frac{N\alpha\beta(\alpha+\beta+N)}{(\alpha+\beta)^2(\alpha+\beta+1)}.
		\end{align}

	\subsection{Fock states}
		To obtain properties of the photon statistics, we have to apply a relation between the true photon statistics $(p_n)_{n\in\mathbb N}$ and the counting statistics $(c_k)_{k=0}^N$, see Eq.~(14) in Ref.~\cite{SpeVoA},
		\begin{align}
			c_k=\sum_{n=0}^\infty \frac{N!}{k!(N-k)!}\frac{\partial_y^n\left(e^y-1\right)^k|_{y=0}}{N^n}\, p_n.
		\end{align}
		The $m$ photon Fock state has a photostatistics given by $p_n=\delta_{m,n}$
		Therefore, we get
		\begin{align}
			c_k=\frac{N!}{k!(N-k)!}\frac{\partial_y^m\left(e^y-1\right)^k|_{y=0}}{N^m}.
		\end{align}
		The mean click number $\langle c\rangle$ and the variance $\langle(\Delta c)^2\rangle$ can be obtained from the generating function, cf. Eq.~(\ref{Eq:GenFct}), which is for the $m$-th Fock state
		\begin{align}
			f(x)=\frac{1}{N^m}\partial_y^m\left.\left[x\left(e^y-1\right)+1\right]^N\right|_{y=0}.
		\end{align}
		The corresponding derivatives read as
		\begin{align}
			\langle c\rangle=\partial_xf(x)|_{x=1}=&\frac{N}{N^m}\partial_y^m\left.\left(e^{Ny}-e^{(N-1)y}\right)\right|_{y=0}
			=N\left(1-\left[1-\frac{1}{N}\right]^m\right),\\
			\langle c^2\rangle-\langle c\rangle=\partial_x^2f(x)|_{x=1}=&
			\frac{N(N-1)}{N^m}\partial_y^m\left.\left(e^y-1\right)^2e^{(N-2)y}\right|_{y=0}
			=N(N-1)\left(1-2\left[1-\frac{1}{N}\right]^m+\left[1-\frac{2}{N}\right]^m\right).
		\end{align}
		Note that the derivatives $D_{k,n}=\partial_y^n\left(e^y-1\right)^k|_{y=0}$ can be obtaind from the following simple facts
		\begin{align}
			D_{k,n}=&0 \hspace*{5cm}\text{(for $k>n$)}\\
			D_{k,n+1}=&k\left(D_{k,n}+D_{k-1,n}\right).
		\end{align}

	\subsection{SPATS}
		According to the result of Eq.~(\ref{Eq:QB}), we need for the calculation of $Q_{\rm B}$ for the SPATS with 
		\begin{align}
			P_{\rm SPATS}(\alpha)=\frac{1}{\pi\bar n^3}\left[(1+\bar n)|\alpha|^2-\bar n\right]e^{-\frac{|\alpha|^2}{\bar n}},
		\end{align}
		the following form of integrals
		\begin{align}
			\nonumber I(\lambda)=&\int d^2\alpha\,P_{\rm SPATS}(\alpha)\, e^{-\lambda|\alpha|^2}=\frac{1+\bar n}{\pi\bar n^3}\int d^2\alpha\,|\alpha|^2 e^{-\left(\frac{1}{\bar n}+\lambda\right)|\alpha|^2}-\frac{1}{\pi\bar n^2}\int d^2\alpha\,e^{-\left(\frac{1}{\bar n}+\lambda\right)|\alpha|^2}\\
			\nonumber =&\frac{1+\bar n}{\bar n^3}\int_0^\infty ds\, s\, e^{-\left(\frac{1}{\bar n}+\lambda\right)s}-\frac{1}{\bar n^2}\int_0^\infty ds\, e^{-\left(\frac{1}{\bar n}+\lambda\right)s}=\frac{1+\bar n}{\bar n^3}\frac{1!}{\left(\frac{1}{\bar n}+\lambda\right)^2}-\frac{1}{\bar n^2}\frac{0!}{\left(\frac{1}{\bar n}+\lambda\right)^1}\\
			=&\frac{1-\lambda}{(1+\lambda \bar n)^2}.
		\end{align}
		Thus, we obtain $\langle{:}\exp[-\eta\hat n/N]{:} \rangle=I(\eta/N)$ and
		\begin{align}
			\langle{:}\left(\Delta e^{-\frac{\eta\hat n}{N}}\right)^2{:} \rangle=\langle{:}e^{-2\frac{\eta\hat n}{N}}{:} \rangle-\langle{:}e^{-\frac{\eta\hat n}{N}}{:} \rangle^2=I(2\eta/N)-I(\eta/N)^2.
		\end{align}
\end{widetext}

\end{document}